\documentclass[pre,twocolumn,showpacs,showkeys,superscriptaddress,preprintnumbers,floatfix]{revtex4-1}

\usepackage{etex}
\usepackage{ifpdf}
\usepackage{hyperref}
\usepackage{dcolumn}
\usepackage{url}
\usepackage{amsmath}
\usepackage{amscd}
\usepackage{amsfonts}
\usepackage{amssymb}
\usepackage{amsthm}
\usepackage{xfrac}
\usepackage{braket}
\usepackage{bm}   
\usepackage{bbm}
\usepackage{verbatim}
\usepackage{stmaryrd}
\usepackage{xcolor}
\usepackage{setspace}
\usepackage{tikz}
\usetikzlibrary{arrows}
\usetikzlibrary{automata}
\usetikzlibrary{positioning}
\usetikzlibrary{plotmarks}
\usepackage{pgfplots}
\pgfplotsset{compat=newest}

\tikzstyle{vaucanson}=[
  node distance=2.5cm,
  bend angle=15,
  auto,
  every loop/.style={},
  every edge/.style={->,draw=black,line width=1.2,>=latex,shorten <=1pt,shorten >=1pt},
  every state/.style={draw=black,line width=2,font=\large},
  loop right/.style={right,out=22,in=-22,loop},
  loop above/.style={above,out=112,in=68,loop},
  loop left/.style={left,out=202,in=158,loop},
  loop below/.style={below,out=292,in=248,loop},
  loop above right/.style={right,out=67,in=23,loop},
  loop above left/.style={left,out=157,in=113,loop},
  loop below left/.style={left,out=247,in=203,loop},
  loop below right/.style={right,out=337,in=293,loop},
]

\theoremstyle{plain}    \newtheorem{Lem}{Lemma}
\theoremstyle{plain}    \newtheorem*{ProLem}{Proof}
\theoremstyle{plain}    
\theoremstyle{plain}    
\theoremstyle{plain}    \newtheorem{The}{Theorem}
\theoremstyle{plain}    \newtheorem*{ProThe}{Proof}
\theoremstyle{plain}    
\theoremstyle{plain}    
\theoremstyle{plain}    
\theoremstyle{plain}    \newtheorem*{Rem}{Remark}
\theoremstyle{plain}    \newtheorem{Def}{Definition}
\theoremstyle{plain}    
\theoremstyle{plain}    

\newcommand{\eM}     {\mbox{$\epsilon$-machine}}
\newcommand{\eMs}    {\mbox{$\epsilon$-machines}}
\newcommand{\EM}     {\mbox{$\epsilon$-Machine}}

\newcommand{\MeasAlphabet}  {\mathcal{A}}
\newcommand{\MeasTime}   { \mathcal{T} }
\newcommand{\meastime}   { \tau }
\newcommand{\MeasSymbol}   { X }
\newcommand{\meassymbol}   { x }
\newcommand{\MeasSymbolTime}   { (X,\mathcal{T}) }
\newcommand{\meassymboltime}   { (x,{\tau}) }

\newcommand{\BiInfinity}    { \smash{\overleftrightarrow {\MeasTime}} }
\newcommand{\biinfinity}    { \smash{\overleftrightarrow {\meastime}} }
\newcommand{\Past} { \MeasTime_{:0^+} }
\newcommand{\past} { \meastime_{:0^+} }
\newcommand{\Future} { \MeasTime_{0^-:} }
\newcommand{\future} { \meastime_{0^-:} }
\newcommand{\pastprime} { \smash{{\past}^{\prime}}}
\newcommand{\futureprime}   { \smash{\future^{\prime}} }

\newcommand{\CausalState}   { \mathcal{S} }

\newcommand{\causalstate}   { \sigma }
\newcommand{\CausalStateSet}    { \boldsymbol{\CausalState} }
\newcommand{\AlternateState}    { \mathcal{R} }

\newcommand{\AlternateStateSet} { \boldsymbol{\AlternateState} }

\newcommand{\Prob}      {\Pr} 

\newcommand{\Cmu}       {C_\mu}
\newcommand{\hmu}       {h_\mu}
\newcommand{\EE}        {{\bf E}}

\newcommand{\ProcessAlphabet}   {\MeasAlphabet}

\newcommand{\forward}{+}
\newcommand{\reverse}{-}
\newcommand{\forwardreverse}{\pm} 

\newcommand{\FutureCausalState} { {\CausalState}^{\forward} }

\newcommand{\PastCausalState}   { {\CausalState}^{\reverse} }

\newcommand{\lastindex}[2]{
  \edef\tempa{0}
  \edef\tempb{#2}
  \ifx\tempa\tempb
    \edef\tempc{#1}
  \else
    \edef\tempa{0}
    \edef\tempb{#1}
    \ifx\tempa\tempb
      \edef\tempc{#2}
    \else
      \edef\tempc{#1+#2}
    \fi
  \fi
  \tempc
}

\newcommand{\rhomu}{\rho_\mu}
\newcommand{\rmu}{r_\mu}
\newcommand{\bmu}{b_\mu}
\newcommand{\qmu}{q_\mu}

\newcommand{\I}{\mathbf{I}}

\newcommand{\CSjoint}[1][,]{
   \edef\tempa{:}
   \edef\tempb{#1}
   \ifx\tempa\tempb
      \ensuremath{\FutureCausalState\!#1\PastCausalState}
   \else
      \ensuremath{\FutureCausalState#1\PastCausalState}
   \fi
}

\newif\ifpm
\edef\tempa{\forwardreverse}
\edef\tempb{\pm}
\ifx\tempa\tempb
   \pmfalse
\else
   \pmtrue
\fi

\newcommand{\CS}{\CausalState}
\newcommand{\cs}{\causalstate}

\renewcommand{\H}{\operatorname{H}}
\renewcommand{\I}{\operatorname{I}}

\colorlet {R_color}    {blue}
\colorlet {k_color}    {black!30!green}

\def\clap#1{\hbox to 0pt{\hss#1\hss}}

\begin{document}

\title{Informational and Causal Architecture of\\
Continuous-time Renewal and Hidden Semi-Markov Processes}

\author{Sarah Marzen}
\email{semarzen@mit.edu}
\affiliation{Physics of Living Systems Group, Department of Physics, Massachusetts Institute of Technology, Cambridge, MA 02139} \affiliation{Department of Physics, University of California at Berkeley, Berkeley, CA 94720}

\author{James P. Crutchfield}
\email{chaos@ucdavis.edu}
\affiliation{Complexity Sciences Center and Department of Physics, University of California at Davis, One Shields Avenue, Davis, CA 95616}

\date{\today}
\bibliographystyle{unsrt}

\begin{abstract}
We introduce the minimal maximally predictive models (\eMs) of processes
generated by certain hidden semi-Markov models. Their causal states are either
hybrid discrete-continuous or continuous random variables and causal-state
transitions are described by partial differential equations. Closed-form
expressions are given for statistical complexities, excess entropies, and
differential information anatomy rates. We present a complete analysis of the
\eMs\ of continuous-time renewal processes and, then, extend this to processes
generated by unifilar hidden semi-Markov models and semi-Markov models. Our
information-theoretic analysis leads to new expressions for the entropy rate
and the rates of related information measures for these very general
continuous-time process classes.
\end{abstract}


\preprint{Santa Fe Institute Working Paper 16-10-XXX}
\preprint{arxiv.org:1610.XXXXX [physics.gen-ph]}

\maketitle


\setstretch{1.1}

\newcommand{\Abet}{\ProcessAlphabet}
\newcommand{\MST}{\MeasSymbolTime}
\newcommand{\mst}{\meassymboltime}
\newcommand{\MT}{\MeasTime}
\newcommand{\mt}{\meastime}
\newcommand{\MS}{\MeasSymbol}
\newcommand{\ms}{\meassymbol}
\newcommand{\SSet}{\CausalStateSet}
\newcommand{\St}{\CausalState}
\newcommand{\st}{\causalstate}
\newcommand{\MxSt}{\AlternateState}
\newcommand{\MxSSet}{\AlternateStateSet}
\newcommand{\mxst}{\mu}
\newcommand{\mxstt}[1]{\mu_{#1}}
\newcommand{\Gen}{\mathcal{G}}
\newcommand{\StartMS}{\bra{\delta_\pi}}


\vspace{0.2in}
\section{Introduction}

We are interested in answering two very basic questions about continuous-time,
discrete-symbol stochastic processes:
\begin{itemize}
\item What are their minimal maximally predictive models---their \eMs?
\item What are information-theoretic characterizations of their randomness, predictability, and complexity?
\end{itemize}
For shorthand, we refer to the former as \emph{causal architecture} and the
latter as \emph{informational architecture}. Minimal maximally predictive
models of discrete-time, discrete-state, discrete-output processes are
relatively well understood; e.g., see Refs. \cite{Shal98a,Lohr09a,Crut13a}.
Some progress has been made on understanding minimal maximally predictive
models of discrete-time, continuous-output processes; e.g., see Refs.
\cite{Kell12a,Marz14a,Riec16a}. Relatively less is understood about minimal
maximally predictive models of continuous-time, discrete-output processes,
beyond those with exponentially decaying state-dwell times \cite{Riec16a}. The
following is a first attempt at a remedy that complements the spectral methods
developed in Ref. \cite{Riec16a}, as we address the less tractable case of
uncountably infinite causal states.

We start by analyzing continuous-time renewal processes, as addressing the
challenges there carries over to other continuous-time processes. (Elsewhere,
we outline the wide interest and applicability of renewal processes in physics
and the quantitative sciences generally \cite{Marz14b,Marzen20161517,Marz14e}.)
The difficulties are both technical and conceptual. First, the causal states
are now continuous or hybrid discrete-continuous random variables, unless the
renewal process is Poisson. Second, transitions between causal states are now
described by partial differential equations. Finally, and perhaps most
challenging, most informational architecture quantities must be redefined. With
these challenges addressed, we turn our attention to a very general class of
continuous-time, discrete-alphabet processes---stateful renewal processes
generated by unifilar hidden semi-Markov models. We identify their \eMs\ and
find new expressions for entropy rate and other informational architecture
quantities, extending results in Ref. \cite{girardin2006entropy}.

Our main thesis is rather simple: minimal maximally predictive models of
continuous-time, discrete-symbol processes require a wholly new \eM\ calculus.
To develop it, Sec.~\ref{sec:Background} describes the required new notation
and definitions that enable extending the \eM\ framework which is otherwise
well understood for discrete-time processes \cite{Shal98a,Crut01a}.
Sections~\ref{sec:CausalStates}-\ref{sec:InfoArch} determine the causal and
informational architecture of continuous-time renewal processes.
Section~\ref{sec:UHSMMs} characterizes the \eMs\ and calculates the entropy
rate and excess entropy of unifilar hidden semi-Markov models. We conclude by
describing potential applications to Bayesian \eM\ inference algorithms using
new enumerations of \eM\ topologies and to information measure estimation using
the formulae of Sec.~\ref{sec:UHSMMs}.


\section{Background and Notation}
\label{sec:Background}

A continuous-time, discrete-symbol time series $\ldots, (\ms_{-1},\mt_{-1}),
(\ms_0,\mt_0),(\ms_1,\mt_1),\ldots$ is described by a list of symbols $\ms$ in
a finite alphabet $\Abet$ and dwell times $\mt \in \Re_{\geq 0}$ for those
symbols.  In this representation, we demand that $\ms_i \neq \ms_{i+1}$ to enforce a unique
presentation of the time series.

Sections~\ref{sec:CausalStates}-\ref{sec:InfoArch} focus on point processes
for which $|\Abet|=1$. And so, in this case, we label the time series only with
dwell times: $\ldots, \mt_{-1}, \mt_0, \mt_1,\ldots$. We view the time series
$\biinfinity$ as a realization of random variables $\BiInfinity = \ldots,
\MT_{-1}, \MT_0, \MT_1,\ldots$. When the observed time series is strictly
stationary and the process ergodic, in principle, we can calculate the
probability distribution $\Prob(\BiInfinity)$ from a single realization $\biinfinity$.

Demarcating the present splits $\MT_0$ into two parts: the time $\MT_{0^+}$
since first emitting the previous symbol and the time $\MT_{0^-}$ to next
symbol. Thus, we define $\MeasTime_{-\infty:0^+} = \ldots, \MT_{-1},\MT_{0^+}$
as the \emph{past} and $\MeasTime_{0^-:\infty} = \MT_{0^-}, \MT_{1},\ldots$ as
the \emph{future}. (To reduce notation, we drop the $\infty$ indices.) The
\emph{present} $\MT_{0^+:0^-}$ itself extends over an infinitesimally small
length of time.

\begin{figure}
\centering
\includegraphics[width=0.5\columnwidth]{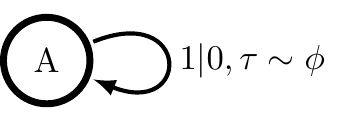}
\caption{Generative model of a continuous-time renewal process. The length
	$\MT_i = \tau$ of periods of silence (corresponding to output symbol $0$)
	are drawn independently, identically distributed (IID) from probability
	density $\phi(t)$.
	}
\label{fig:genModel}
\end{figure}

Continuous-time renewal processes have a relatively simple generative model.
\emph{Interevent intervals} $\MT_i$ are drawn from a probability density
function $\phi(t)$. The \emph{survival function} $\Phi(t) =
\int_t^{\infty}\phi(t') dt'$ is the probability that an interevent interval is
greater than or equal to $t$ and, in a nod to neuroscience, we define the
\emph{mean firing rate} $\mu$ as:
\begin{align*}
\mu^{-1} = \int_0^{\infty} t\phi(t) dt
  ~.
\end{align*}
The minimal generative model for a continuous-time renewal process is therefore
a single causal-state machine with a continuous-value observable $\MT$; as shown
in Fig.~\ref{fig:genModel}.

\subsection{Causal architecture}

A process' \emph{forward-time causal states} are defined, as usual, by the
\emph{predictive} equivalence relation \cite{Shal98a}, written here for the case of point processes:
\begin{align*}
\past \sim_{\epsilon^+}& \pastprime \\
& \Leftrightarrow \Prob(\Future|\Past=\past)=\Prob(\Future|\Past=\pastprime)
  ~.
\end{align*}
It is straightforward to write the predictive equivalence relation for continuous-time, discrete-alphabet point processes using the notation.
This partitions the set of allowed pasts. Each equivalence class of pasts
is a forward-time causal state $\st^+ = \epsilon^+(\past)$, in which
$\epsilon^+(\cdot)$ is the function that maps a past to its causal state.  The
set of forward-time causal states $\St^+ = \{\st^+\}$ inherits a probability
distribution $\Prob(\St^+)$ from the probability distribution over pasts
$\Prob(\Past)$. \emph{Forward-time prescient statistics} are any refinement of
the forward-time causal-state partition. By construction, they are a sufficient
statistic for prediction, but not necessarily \emph{minimal} sufficient
statistics \cite{Shal98a}.

\emph{Reverse-time causal states} are essentially forward-time causal states of
the time-reversed process. In short, reverse-time causal states
$\St^-=\{\st^-\}$ are the classes defined by the retrodictive equivalence
relation, written here for the case of point processes:
\begin{align*}
\future \sim_{\epsilon^-}& \futureprime \\
& \Leftrightarrow\Prob(\Past|\Future=\future)=\Prob(\Past|\Future=\futureprime)
  ~.
\end{align*}
It is, again, straightforward to write the predictive equivalence relation for
continuous-time, discrete-alphabet point processes using the notation given
above. And, similarly, reverse-time causal states $\St^- = \epsilon^-(\Future)$
inherit a probability measure $\Prob(\St^-)$ from the probability distribution
$\Prob(\Future)$ over futures. Reverse-time prescient statistics are any
refinement of the reverse-time causal-state partition. They are sufficient
statistics for retrodiction, but not necessarily minimal.

\begin{figure}[htp]
\centering
\includegraphics[width=\columnwidth]{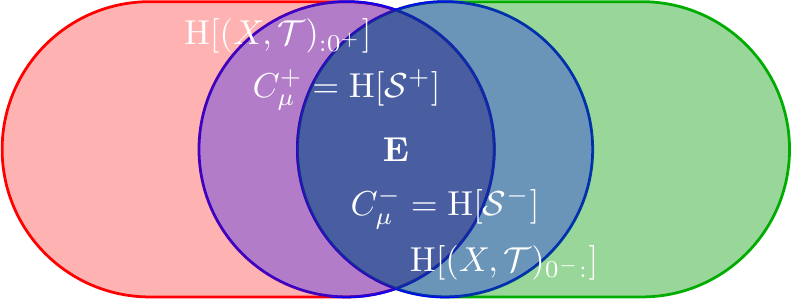}
\caption{Predictability, compressibility, and causal irreversibility in
	renewal and semi-Markov processes graphically illustrated using a Venn-like
	\emph{information diagram} over the random variables for the past
	$(\MS,\MT)_{:0^+}$ (left oval, red), the future $(\MS,\MT)_{0^-:}$ (right
	oval, green), the forward-time causal states $\St^+$ (left circle, purple),
	and the reverse-time causal states $\St^-$ (right circle, blue). (Cf. Ref.
	\cite{Crut08b}.) The forward-time and reverse-time statistical complexities
	are the entropies of $\St^+$ and $\St^-$, i.e., the memories required to
	losslessly predict or retrodict, respectively. The excess entropy $\EE =
	\I[(\MS,\MT)_{:0^+};(\MS,\MT)_{0^-:}]$ is a measure of process
	predictability (central pointed ellipse, dark blue) and Theorem $1$ of Ref.
	\cite{Crut08a,Crut08b} shows that $\EE = \I[\St^+;\St^-]$ by applying the
	causal shielding relations in Eqs.~(\ref{eq:CS1}) and (\ref{eq:CS2}).
  }
\label{fig:Intuition2}
\end{figure}

The main import of these definitions derives from the \emph{causal shielding}
relations:
\begin{align}
\Prob(\Future,\Past|\St^+) & = \Prob(\Future|\St^+) \Prob(\Past|\St^+) 
  \label{eq:CS1} \\
\Prob(\Future,\Past|\St^-) & = \Prob(\Future|\St^-) \Prob(\Past|\St^-)
  ~.
\label{eq:CS2}
\end{align}
The consequence of these is illustrated in Fig.~\ref{fig:Intuition2}. That is,
arbitrary functions of the past and future do not shield the two aggregate past
and future random variables from one another. So, these causal shielding
relations are special to prescient statistics, causal states, and their
defining functions $\epsilon^+(\cdot)$ and $\epsilon^-(\cdot)$. Forward and
reverse-time generative models do not, in general, have state spaces that
satisfy Eqs.~(\ref{eq:CS1}) and (\ref{eq:CS2}).

The \emph{forward-time \eM} is that with state space $\St^+$ and transition
dynamic between forward-time causal states. The \emph{reverse-time \eM} is that
with state space $\St^-$ and transition dynamic between reverse-time causal
states. Defining these transition dynamics for continuous-time processes
requires a surprising amount of care, as discussed in
Secs.~\ref{sec:CausalStates}-\ref{sec:InfoArch}.

\subsection{Informational architecture}

We are broadly interested in information-theoretic characterizations of a
process' predictability, compressibility, and randomness.  A list of current
quantities of interest, though by no means exhaustive, is given in
Figs.~\ref{fig:Intuition2} and \ref{fig:Intuition3}. Curiously, many lose
meaning when naively applied to continuous-time processes; e.g., see Refs.
\cite{Gasp93a,Marz14a,Marz14e}. This section, as a necessity, will redefine
many of these in relatively simple, but new ways to avoid trivial divergences
and zeros.

The \emph{forward-time statistical complexity} $\Cmu^+ = \H[\St^+]$ is the cost
of coding the forward-time causal states and the \emph{reverse-time statistical
complexity} $\Cmu^- = \H[\St^-]$ is the cost of coding reverse-time causal
states. When $\St^+$ or $\St^-$ are mixed or continuous random variables, one
employs differential entropies for $\H[\cdot]$. The result, though, is that the
statistical complexities are potentially negative or infinite or both
\cite[Ch.  8.3]{Cove91a}, perhaps undesirable characteristics for a definition
of process complexity. This definition, however, allows for consistency with
complexity definitions for discretized continuous-time processes. See Ref.
\cite{rao2004cumulative} for possible alternatives for $\H[\cdot]$.

Together, a process' \emph{causal irreversibility} \cite{Crut08a,Crut08b} is defined as the difference between the forward and reverse-time statistical complexities:
\begin{align*}
\Xi = \Cmu^+ - \Cmu^-
  ~.
\end{align*}
If the forward- and reverse-time process \eMs\ are isomorphic---i.e., if the
process is temporally reversible---then $\Xi = 0$.

\newcommand{\PresentP}{\Gamma_{\delta}}
\newcommand{\FutureP}{\MT_{\delta^-:}}

\section{Continuous-time causal states}
\label{sec:CausalStates}

Renewal processes are temporally symmetric: $\Xi = 0$ \cite{Marz14b}. As such,
we will refer to forward-time causal states and the forward-time \eM\ as simply
causal states or the \eM, with the understanding that reverse-time causal
states and reverse-time \eMs\ will take the exact same form with slight
labeling differences.

We start by describing prescient statistics for continuous-time processes. The
Lemma which does this exactly parallels that of Lemma $1$ of Ref.
\cite{Marz14b}. The only difference is that the prescient statistic is the
\emph{time} since last event, rather than the number of $0$s (count) since last event.

{\Lem The time $\MT_{0^+}$ since last event is a prescient statistic of renewal
processes.
\label{lem:ctrp_prescient}
}

{\ProLem From Bayes Rule:
\begin{align*}
\Prob(\MT_{0^-:}|\MT_{:0^+})
  & = \Prob(\MT_{0^-}|\MT_{0^+:})\Prob(\MT_{1:}|\MT_{:1})
  ~.
\end{align*}
Interevent intervals $\MT_i$ are independent of one another, so
$\Prob(\MT_{1:}|\MT_{:1}) = \Prob(\MT_{1:})$. The random variables $\MT_{0^+}$
and $\MT_{0^-}$ are functions of $\MT_0$ and the location of the present. Both
$\MT_{0^+}$ and $\MT_{0^-}$ are independent of other interevent intervals. And
so, $\Prob(\MT_{0^-}|\MT_{0^+:}) = \Prob(\MT_{0^-}|\MT_{0^+})$. This implies:
\begin{align}
\Prob(\MT_{0^-:}|\MT_{:0^+}) = \Prob(\MT_{1:})\Prob(\MT_{0^-}|\MT_{0^+})
\label{eq:LemProof}
  ~.
\end{align}
The predictive equivalence relation groups two pasts $\mt_{:0^+}$ and
$\mt_{:0^+}'$ together when $\Prob(\MT_{0^-:}|\MT_{:0^+}=\mt_{:0^+}) =
\Prob(\MT_{0^-:}|\MT_{:0^+}=\mt_{:0^+}')$.  We see that $\mt_{0^+} =
\mt_{0^+}'$ is a sufficient condition for this from Eq.~(\ref{eq:LemProof}).
The Lemma follows.
}

Some renewal processes are quite predictable, while others are purely random.
A Poisson process is the latter: Interevent intervals are drawn independently
from an exponential distribution and so knowing the time since last event
provides no predictive benefit. A fractal renewal process can be the former.
There, the interevent interval is so structured that the resultant process can
have power-law correlations \cite{Lowe93a}. Then, knowing the time since last
event can provide quite a bit of predictive power \cite{Marzen20161517}.

Intermediate between these two extremes is a broad class of renewal processes
whose interevent intervals are structured up to a point and then fall off
exponentially only after some time $T^*$. These intermediate cases can be
classified as either of the following types of renewal process, in analogy with
Ref. \cite{Marz14b}'s classification.  Note that an eventually $\Delta$-Poisson
process, but not an eventually Poisson process, will generally have a discontinuous $\phi(t)$.

{\Def An \emph{eventually Poisson process} has:
\begin{align*}
\phi(t) = \phi(T) e^{-\lambda (t-T)}
  ~,
\end{align*}
for some $\lambda>0$ and $T>0$ almost everywhere. We associate the eventually Poisson process with the minimal such $T$.
\label{def:EP}
}

{\Def An \emph{eventually $\Delta$-Poisson process} with $\Delta^* > 0$ has an
interevent interval distribution satisfying:
\begin{align*}
\phi(T^*+s)
  = \phi(T^*+(s-T^*) \!\!\!\!\mod \Delta^*) e^{-\lambda\lfloor s/\Delta^*\rfloor}
\end{align*}
for the smallest possible $T^*$ for which $\Delta^*$ exists.
\label{def:EDP}
}

A familiar example of an eventually Poisson process is found in the spike
trains generated by Poisson neurons with refractory periods \cite{Marz14e}.
There, the neuron is effectively prevented from firing two spikes within a time
$T$ of each other---the period during with its ion channels re-energize the
membrane voltage to their nonequilibrium steady state. After that, the time to
next spike is drawn from an exponential distribution. To exactly predict the
spike train's future, we must know the time since last spike, as long as it is
less than $T$. We gain a great deal of predictive power from that piece of
information. However, we do not care much about the time since last spike
exactly if it is greater than $T$, since at that point the neuron acts as a
memoryless Poisson neuron. These intuitions are captured by the following
classification theorem.

{\The A renewal process has three different types of causal state:
\begin{enumerate}
\item When the renewal process is not eventually $\Delta$-Poisson, the causal states are the time since last event;
\item When the renewal process is eventually Poisson, the causal states are the
time since last event up until time $T^*$; or
\item When the renewal process is eventually $\Delta$-Poisson, the causal
states are the time since last event up until time $T^*$ and are the times
since $T^*$ mod $\Delta$ thereafter.
\end{enumerate}
\label{the:RenewCausalStates}
}

{\ProThe Lemma \ref{lem:ctrp_prescient} implies that two pasts are causally
equivalent if they have the same time since last event, if
$\mt_{0^+}=\mt_{0^+}'$. From Lemma \ref{lem:ctrp_prescient}'s proof, we further
see that two times since last event are causally equivalent when
$\Prob(\MT_{0^-}|\MT_{0^+}=\mt_{0^+}) = \Prob(\MT_{0^-}|\MT_{0^+}=\mt_{0^+}')$.
In terms of $\phi(t)$, we find that:
\begin{align*}
\Prob(\MT_{0^-} = \mt_{0^-}|\MT_{0^+}=\mt_{0^+})
  = \frac{\phi(\mt_{0^-}+\mt_{0^+})}{\Phi(\mt_{0^+})}
  ~,
\end{align*}
using manipulations very similar to those in the proof of Thm.~$1$ of Ref.
\cite{Marz14b}. So, to find causal states, we look for $\mt_{0^+}\neq
\mt_{0^+}'$ such that:
\begin{align*}
\frac{\phi(\mt_{0^-}+\mt_{0^+})}{\Phi(\mt_{0^+})} = \frac{\phi(\mt_{0^-}+\mt_{0^+}')}{\Phi(\mt_{0^+}')}
  ~.
\end{align*}
for all $\mt_{0^-}\geq 0$.

To unravel the consequences of this, we suppose that $\mt_{0^+}< \mt_{0^+}'$
without loss of generality. Define $\Delta = \mt_{0^+}'-\mt_{0^+}$ and $T =
\mt_{0^+}$, for convenience. The predictive equivalence relation can then be
rewritten as:
\begin{align*}
\phi(T+\Delta+\mt_{0^-}) = \lambda \phi(T+\mt_{0^-})
  ~,
\end{align*}
for any $\mt_{0^-}\geq 0$, where $\lambda = \Phi(T+\Delta) / \Phi(T)$.
Iterating this relationship, we find that:
\begin{align*}
\phi(T+\mt_{0^-})
  = \lambda^{\lfloor \mt_{0^-}/\Delta \rfloor}
  \phi \left( T+ (\mt_{0^-} \!\!\!\! \mod \Delta) \right)
  ~.
\end{align*}
This immediately implies the theorem's first case. If a renewal process is
\emph{not} eventually $\Delta$-Poisson, then $\phi(\mt_{0^-}+\mt_{0^+}) /
\Phi(\mt_{0^+}) = \phi(\mt_{0^-}+\mt_{0^+}') / \Phi(\mt_{0^+}')$ for all
$\mt_{0^-}\geq 0$ implies $\mt_{0^+}=\mt_{0^+}'$, so that the prescient
statistics of Lemma \ref{lem:ctrp_prescient} are also minimal.

To understand the theorem's last two cases, we consider more carefully the set
of all pairs $(T,\Delta)$ for which $\phi(\mt_{0^-}+T) / \Phi(T) =
\phi(\mt_{0^-}+T+\Delta) / \Phi(T+\Delta)$ for all $\mt_{0^-}\geq 0$ holds.
Define the set:
\begin{align*}
\CS_{T,\Delta} &:= \big\{ (T,\Delta): \\
  & \frac{\phi(\mt_{0^-}+T)}{\Phi(T)}
  = \frac{\phi(\mt_{0^-}+T+\Delta)}{\Phi(T+\Delta)},
  \text{~for~all~} \mt_{0^-} \geq 0 \big\} 
\end{align*}
and define the parameters $T^*$ and $\Delta^*$ by:
\begin{align*}
T^* & := \inf \{T: ~\text{there~exists}~
  \Delta ~\text{such that }~ (T,\Delta) \in \CS_{T,\Delta} \}
\end{align*}
and:
\begin{align*}
  \Delta^* & := \inf \{\Delta : (T^*,\Delta)\in \CS_{T,\Delta}\}
  ~.
\end{align*}
Note that $T^*$ and $\Delta^*$ defined in this way are unique and exist, as we
assumed that $\CS_{T,\Delta}$ is nonempty. When $\Delta^*>0$, then the
process is eventually $\Delta$-Poisson.  If $\Delta^*=0$, then the process must
be an eventually Poisson process with parameter $T^*$. To see this, we return
to the equation:
\begin{align*}
\phi(T^*+\Delta+\mt_{0^-}) = \frac{\Phi(T^*+\Delta)}{\Phi(T^*)} \phi(T^*+\mt_{0^-})
  ~,
\end{align*}
and rearrange terms to find:
\begin{align*}
\frac{\phi(T^*+\Delta+\mt_{0^-})-\phi(T^*+\mt_{0^-})}{\phi(T^*+\mt_{0^-})} = \frac{\Phi(T^*+\Delta)-\Phi(T^*)}{\Phi(T^*)}.
\end{align*}
As $\Delta^* = 0$, we can take the limit that $\Delta\rightarrow 0$ and we find that:
\begin{align*}
\frac{d\log\phi(t)}{dt} \big|_{t=T^*+\mt_{0^-}}
  = \frac{d\log\Phi(t)}{dt} \big|_{t=T^*}
  ~.
\end{align*}
The righthand side is a parameter independent of $\mt_{0^-}$. So, this is a
standard ordinary differential equation for $\phi(t)$. It is solved by $\phi(t)
= \phi(T^*) e^{-\lambda (t-T^*)}$ for $\lambda := -d\log\Phi(t) / dt
\big|_{t=T^*}$.
}

Theorem~\ref{the:RenewCausalStates} implies that there is a qualitative change
in $\St^+$ depending on whether or not the renewal process is Poisson,
eventually Poisson, eventually $\Delta$-Poisson, or not eventually Poisson. In
the first case, $\St^+$ is a discrete random variable; in the second case,
$\St^+$ is a mixed discrete-continuous random variable; and in the third and
fourth cases, $\St^+$ is a continuous random variable.

\section{Wave propagation on Continuous-time \eMs}

Identifying causal states in continuous-time follows an almost entirely similar
path to that used for discrete-time renewal processes in Ref. \cite{Marz14b}.
The seemingly slight differences between the causal states of eventually
Poisson, eventually $\Delta$-Poisson, and not eventually $\Delta$-Poisson
renewal processes, however, have surprisingly important consequences for
continuous-time \eMs.

As described by Thm.~\ref{the:RenewCausalStates}, there are often an
uncountable infinity of continuous-time causal states. As one might anticipate
from Refs. \cite{Marz14b,Marz14e}, however, there is an ordering to this
infinity of causal states that makes calculations tractable. There is one major
difference between discrete-time \eMs\ and continuous-time \eMs: transition
dynamics often amount to specifying the evolution of a probability density
function over causal-state space.

As such, a continuous-time \eM\ constitutes an unusual presentation of a hidden
Markov model: they appear as a system of conveyor belts or, under special
conditions, like conveyor belts with a trash bin or a second mini-conveyor
belt. Beyond the picaresque metaphor, in fact they operate like conveyor belts
in that they transport the time since the last event, resetting it here and
there in a stateful way.

Unsurprisingly, the exception to this general rule is given by the Poisson
process itself. The \eM\ of a Poisson process is exactly the minimal generative
model shown in Fig.~\ref{fig:genModel}.  At each iteration, an interevent
interval is drawn from a probability density function $\phi(t) = \lambda
e^{-\lambda t}$, with $\lambda >0$.  Knowing the time since last event does not
aid in predicting the time to next event, above and beyond knowing $\lambda$.
And so, the Poisson \eM\ has only a single state.

In the general setting, though, the \eM\ dynamic describes the evolution of the
probability density function over its causal states. How to represent this? We
might search for labeled transition operators $\mathcal{O}^{(x)}$ such that
$\partial\rho(\st,t) / \partial t = \mathcal{O}^{(x)} \rho(\st,t)$, giving partial differential equations
that govern the labeled-transition dynamics.

\begin{figure}
\includegraphics[width=\columnwidth]{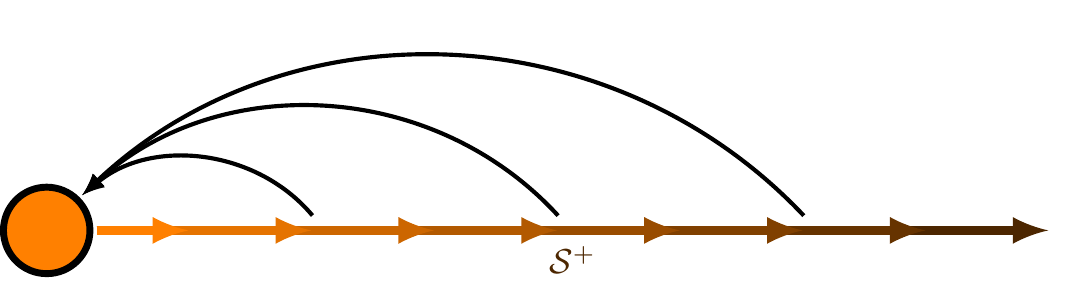}
\caption{\EM\ for the generic not eventually Poisson renewal process:
	Continuous-time causal states $\CS^+$, tracking the time since last event
	and depicted as the semi-infinite horizontal line, are isomorphic with the
	positive real line. If no event is seen, probability flows towards
	increasing time since last event, as described in
	Eq.~(\ref{eq:mathcalO0}). Otherwise, arrows denote allowed transitions
	back to the reset state or ``$0$ node'' (solid black circle at left),
	denoting that an event occurred.
	}
\label{fig:NEDP}
\end{figure}

\subsection{Not Eventually Poisson}
\label{sec:NEDP}

The \eM\ of a renewal process that is not eventually Poisson takes the
state-transition form shown in Fig.~\ref{fig:NEDP}. Let $\rho(\st,t)$ be the
probability density function over the causal states $\st$ at time $t$.  Our
approach to deriving labeled transition dynamics parallels well-known
approaches to determining Fokker-Planck equations using a Kramers-Moyal
expansion \cite{Risk12a}. Here, this means that any probability at causal state
$\st$ at time $t+\Delta t$ could only have come from causal state $\st-\Delta
t$ at time $t$, if $\st\geq \Delta t$. This implies:
\begin{align}
\rho(\st, & t+\Delta t) \nonumber \\
  & = \Prob(\St_{t+\Delta t} = \st|\St_t = \st-\Delta t) \rho(\st-\Delta t,t)
  ~.
\label{eq:ProbFlow1}
\end{align}
However, $\Prob(\St_{t+\Delta t} = \st|\St_t = \st-\Delta t)$ is simply the
probability that the interevent interval is greater than $\st$, given that the
interevent interval is at least $\st-\Delta t$, or:
\begin{align}
\Prob(\St_{t+\Delta t} = \st|\St_t = \st-\Delta t) = \frac{\Phi(\st)}{\Phi(\st-\Delta t)}
  ~.
\label{eq:ProbFlow2}
\end{align}
Together, Eqs. (\ref{eq:ProbFlow1}) and (\ref{eq:ProbFlow2}) imply that:
\begin{align*}
\rho(\st,t+\Delta t) = \frac{\Phi(\st)}{\Phi(\st-\Delta t)} \rho(\st-\Delta t,t)
  ~.
\end{align*}
From this, we obtain:
\begin{align}
\frac{\partial\rho(\st,t)}{\partial t}
  & = \lim_{\Delta t\rightarrow 0} \frac{\rho(\st,t+\Delta t)-\rho(\st,t)}{\Delta t}
  \nonumber \\
  & = \lim_{\Delta t\rightarrow 0} \frac{\frac{\Phi(\st)}{\Phi(\st-\Delta t)} \rho(\st-\Delta t,t)-\rho(\st,t)}{\Delta t}
  \nonumber \\
  & = \lim_{\Delta t\rightarrow 0} \frac{(\frac{\Phi(\st)}{\Phi(\st-\Delta t)} -1)\rho(\st-\Delta t,t)}{\Delta t} \nonumber \\
  & \qquad + \lim_{\Delta t\rightarrow 0} \frac{\rho(\st-\Delta t,t)-\rho(\st,t)}{\Delta t}
  \nonumber \\
  & = \frac{\partial \log \Phi(\st)}{\partial\st} \rho(\st,t)
  - \frac{\partial\rho(\st,t)}{\partial \st}
  ~.
\label{eq:mathcalO0}
\end{align}
Hence, the labeled transition operator $\mathcal{O}^{(0)}$ given no event
takes the form:
\begin{align*}
\mathcal{O}^{(0)} = \frac{\partial \log \Phi(\st)}{\partial\st}
  - \frac{\partial}{\partial\st}
  ~.
\end{align*}
The probability density function $\rho(\st,t)$ changes discontinuously after an
event occurs, though. All probability mass shifts from $\st>0$ resetting back
to $\st=0$:
\begin{align*}
\mathcal{O}^{(1)} \rho(\st,t) = -\frac{\phi(\st)}{\Phi(\st)}\rho(\st,t) + \delta(\st) \int_0^{\infty} \frac{\phi(\st')}{\Phi(\st')} \rho(\st',t) d\st'
  ~.
\end{align*}
In other words, an event ``collapses the wavefunction''.

The stationary distribution $\rho(\st)$ over causal states is given by setting
$\partial_t \rho(\st,t)$ to $0$ and solving. (At the risk of notational
confusion, we adopt the convention that $\rho(\st)$ denotes the stationary
distribution and that $\rho(\st,t)$ does not.) Straightforward algebra shows
that:
\begin{align*}
\rho(\st) = \mu\Phi(\st)
  ~.
\end{align*}

From this, the continuous-time statistical complexity directly follows:
\begin{align*}
\Cmu = \int_0^{\infty} \mu \Phi(\st) \log \frac{1}{\mu\Phi(\st)} d\st
  ~.
\end{align*}
This was the nondivergent component of the infinitesimal time-discretized
renewal process' statistical complexity found in Ref. \cite{Marz14e}.

\begin{figure}
\includegraphics[width=\columnwidth]{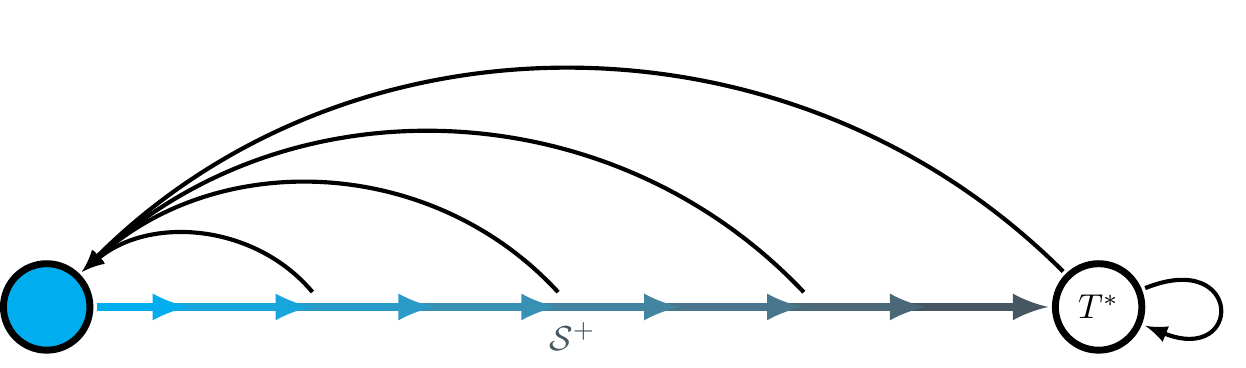}
\caption{\EM\ for an eventually Poisson renewal process: Continuous-time causal
	states $\CS^+$ are isomorphic with the real line only to $[0,T^*]$, as they
	again denote time since last event. A leaky absorbing node at $T^*$
	(solid white circle at right) corresponds to any time since
	last event after $T^*$. If no event is seen, probability flows
	towards increasing time since last event or the leaky
	absorbing node, as described in Eqs.~(\ref{eq:mathcalO0}) and
	(\ref{eq:mathcalO0b}). When an event occurs the process
	transitions (curved arrows) back to the reset state---node $0$
	(solid black circle at left).
	}
\end{figure}

\subsection{Eventually Poisson}

As Thm.~\ref{the:RenewCausalStates} anticipates, there is a qualitatively
different topology to the \eM\ of an eventually Poisson renewal process,
largely due to the continuous-time causal states being mixed
discrete-continuous random variables. For $\st<T^*$, there is ``wave''
propagation completely analogous to that described in Eq.~(\ref{eq:mathcalO0})
of Sec.~\ref{sec:NEDP}. However, there is a new kind of continuous-time causal
state at $\st=T^*$, which does not have a one-to-one correspondence to the
dwell time.  Instead, it denotes that the dwell time is \emph{at least} some
value; viz., $T^*$. New notation follows accordingly: $\rho(\st,t)$, defined
for $\st< T^*$, denotes a probability density function for $\st<T^*$ and
$\pi(T^*,t)$ denotes the probability of existing in causal state $\cs = T^*$.
Normalization, then, requires that:
\begin{align*}
\int_0^{T^*} \rho(\st,t) d\st + \pi(T^*,t) = 1
  ~.
\end{align*}

The transition dynamics for $\pi(T^*,t)$ are obtained similarly to that for
$\rho(\st,t)$, in that we consider all ways in which probability flows to
$\pi(T^*,t+\Delta t)$ in a short time window $\Delta t$. Probability can flow
from any causal state with $T^*-\Delta t \leq \st<T^*$ or from $\st=T^*$
itself. That is, if no event is observed, we have:
\begin{align*}
\pi(T^*,t+\Delta t)
  & = e^{-\lambda \Delta t}\pi(T^*,t) \nonumber \\
  & \qquad + \int_{0^+}^{\Delta t} \rho(T^*-t',t)
  \frac{\Phi(T^*) e^{-\lambda (\Delta t-t')} }{\Phi(T^*-t')} dt'
  .
\end{align*}
The term $e^{-\lambda \Delta t}\pi(T^*,t)$ corresponds to probability flow from
$\st=T^*$ and the integrand corresponds to probability influx from states
$\st=T^*-t'$ with $0<t'\leq\Delta t$. Assuming differentiability of
$\pi(T^*,t)$ with respect to $t$, we find that:
\begin{align}
\frac{\partial}{\partial t} \pi(T^*,t)
  & = -\lambda \pi(T^*,t) + \rho(T^*,t)
  ~,
\label{eq:mathcalO0b}
\end{align}
where $\rho(T^*,t)$ is shorthand for $\lim_{\st\rightarrow T^*} \rho(\st,t)$.
This implies that the labeled transition operator $\mathcal{O}^{(0)}$ takes a
piecewise form which acts as in Eq.~(\ref{eq:mathcalO0}) for $\st<T^*$ and as
in Eq.~(\ref{eq:mathcalO0b}) for $\st=T^*$. As earlier, observing an event
causes the ``wavefunction collapse'' to a delta distribution at $\st=0$.

The causal-state stationary distribution is determined again by setting
$\partial_t\rho(\st,t)$ and $\partial_t \pi(\st,t)$ to $0$.  Equivalently, one
can use the prescription suggested by Thm.~\ref{the:RenewCausalStates} to
calculate $\pi(T^*)$ via integration of the stationary distribution over the
prescient machine given in Sec.~\ref{sec:NEDP}:
\begin{align*}
\pi(T^*) & = \int_{T^*}^{\infty} \rho(\st) d\st \\
  & = \mu \int_{T^*}^{\infty} \Phi(\st) d\st
  ~.
\end{align*}
If we recall that $\Phi(\st) = \Phi(T^*) e^{-\lambda (t-T^*)}$, we find that:
\begin{align*}
\pi(T^*) = \mu \Phi(T^*)/\lambda
  ~.
\end{align*}
The process' continuous-time statistical complexity---precisely, entropy of
this mixed random variable---is given by:
\begin{align*}
\Cmu = \int_0^{T^*} \mu
  \Phi(\st) \log \frac{1}{\mu \Phi(\st)} d\st
  - \frac{\mu \Phi(T^*)}{\lambda} \log \frac{\mu \Phi(T^*)}{\lambda}
  ~.
\end{align*}
This is the sum of the nondivergent $\Cmu$ component and the rate of divergence
of $\Cmu$ of the infinitesimal time-discretized renewal process \cite{Marz14e}.

\begin{figure}
\includegraphics[width=\columnwidth]{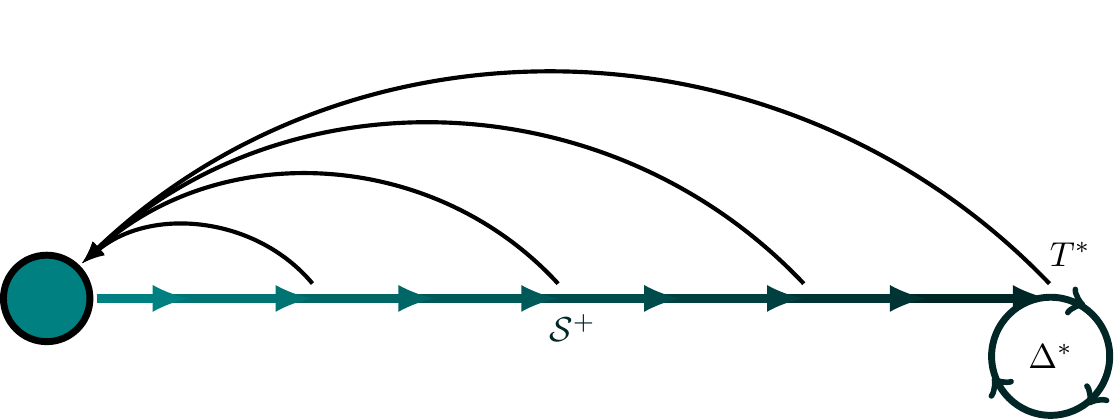}
\caption{\EM\ for an eventually $\Delta$-Poisson renewal process:
	Graphical elements as in the previous figure. The circular causal-state
	space at $T^*$ (circle on right) has total duration $\Delta^*$,
	corresponding to any time since last event after $T^*$ mod $\Delta^*$. If
	no event is seen, probability flows as indicated around the circle, as
	described in Eq.~(\ref{eq:mathcalO0}).
}
\label{fig:EDP}
\end{figure}

\subsection{Eventually-$\Delta$ Poisson}

Probability wave propagation equations, like those in Eq.~(\ref{eq:mathcalO0}),
hold for $\st<T^*$ and for $T^*<\st<T^*+\Delta$. At $\st=T^*$, if no event is
observed, probability flows in from both $(T^*+\Delta)^-$ and from $(T^*)^-$,
giving rise to the equation:
\begin{align*}
\rho(T^*,t+\Delta t) = \rho(T^*-\Delta t,t) + \rho(T^*+\Delta^*-\Delta t,t)
  ~.
\end{align*}
Unfortunately, there is a discontinuous jump in $\rho(\st,t)$ at $\st=T^*$
coming from $(T^*)^-$ and $(T^*+\Delta^*)^-$. And so, we cannot Taylor expand
either $\rho(T^*-\Delta t,t) $ or $\rho(T^*+\Delta^*-\Delta t,t)$ about $\Delta
t=0$.

Again, we can use the prescription suggested by
Thm.~\ref{the:RenewCausalStates} to calculate the probability density function
over these causal states and, from that, calculate the continuous-time
statistical complexity.  Below $\st<T^*$, the probability density function over
causal states is exactly that described in Sec.~\ref{sec:NEDP}: $\rho(\st) =
\mu \Phi(\st)$. For $T^*\leq \st<T^*+\Delta$, the probability density function
becomes:
\begin{align*}
\rho(\st) & = \sum_{\st':(\st'-T^*) \!\!\!\!\!\! \mod\Delta^*=\st}
  \mu \Phi(\st') \\
  & = \mu \sum_{i=0}^{\infty} \Phi(\st+i\Delta^*)
  ~.
\end{align*}
Recalling Def.~\ref{def:EDP}, we see that $\Phi(\st+i\Delta^*) = e^{-\lambda i}
\Phi(\st)$ and so find that for $\st>T^*$:
\begin{align*}
\rho(\st) & = \mu \Phi(\st) \sum_{i=0}^{\infty} e^{-\lambda i} \\
          & = \frac{\mu \Phi(\st)}{1-e^{-\lambda}}
  ~.
\end{align*}
Altogether, this gives the statistical complexity:
\begin{align*}
\Cmu & = \int_0^{T^*}
  \mu \Phi(\st) \log \frac{1}{\mu \Phi(\st)} d\st \nonumber \\
  & \qquad \qquad + \int_{T^*}^{T^*+\Delta^*}
  \frac{\mu \Phi(\st)}{1-e^{-\lambda}}
  \log \frac{1-e^{-\lambda}}{\mu \Phi(\st)} d\st
  ~.
\end{align*}

\begin{table*}[ht] 
\centering
\begin{tabular}{l l}
\hline\hline
\bf Quantity & \bf Expression \\[5pt]
\hline\\[-1.5ex]
$\Cmu^+ = \H[\St^+]$
  & $\int_0^{\infty} \mu \Phi(\st) \log \frac{1}{\mu\Phi(\st)} d\st$
  \hspace{2.4in} Not eventually $\Delta$-Poisson \\[10pt]
  & $ \int_0^{T^*} \mu \Phi(\st) \log \frac{1}{\mu \Phi(\st)} d\st - \frac{\mu \Phi(T^*)}{\lambda} \log \frac{\mu \Phi(T^*)}{\lambda}$
   \hspace{1.56in} Eventually Poisson \\[10pt]
   & $ \int_0^{T^*} \mu \Phi(\st) \log \frac{1}{\mu \Phi(\st)} d\st+
   \int_{T^*}^{T^*+\Delta^*} \frac{\mu \Phi(\st)}{1-e^{-\lambda}} \log
   \frac{1-e^{-\lambda}}{\mu \Phi(\st)} d\st$ \hspace{0.8in} Eventually
   $\Delta$-Poisson \\[10pt]
\hline\\[-1.5ex]
$\EE = \I[\Past;\Future]$ & $\int_0^{\infty} \mu t \phi(t) \log_2
  \big( \mu \phi(t) \big) dt
    -2 \int_0^{\infty} \mu \Phi(t)
  \log_2 \big( \mu \Phi(t) \big) dt$
  \\[5pt]
\hline\\[-1.5ex]
$\hmu = \lim_{\delta\rightarrow 0} \frac{d\H[\PresentP|\Past]}{d\delta}$ &  $-\mu \int_0^{\infty} \phi(t)\log \phi(t) dt$
  \\[5pt]
\hline\\[-1.5ex]
$\bmu = \lim_{\delta\rightarrow 0} \frac{d\I[\FutureP;\PresentP|\Past]}{d\delta}$
  & ~ $-\mu \Big( 2\int_0^{\infty} \phi(t)\log \phi(t+t') dt -1 - \int_0^{\infty} \phi(t) \int_0^{\infty} \phi(t') \log \phi(t+t') dt' dt \Big)$
  \\[5pt]
\hline\\[-1.5ex]
$\qmu =  \lim_{\delta\rightarrow 0} \frac{d\I[\Past;\PresentP;\FutureP]}{d\delta}$
  & ~ $\mu \int_0^{\infty} \phi(t) \int_0^{\infty} \phi(\st^+)\log \phi(t') dt' dt -\mu\log \mu$ \\[5pt]
\hline\\[-1.5ex]
$\rmu = \lim_{\delta\rightarrow 0} \frac{d\H[\PresentP|\Past,\FutureP]}{d\delta}$
  & ~ $ -\mu \Big( 2\int_0^{\infty} \phi(t)\log \phi(t+t') dt -1 - \int_0^{\infty} \phi(t) \int_0^{\infty} \phi(t') \log \phi(t+t') dt' dt \Big)$ \\[5pt]
\hline\\[-1.5ex]
$H_0 = \lim_{\delta\rightarrow 0} \frac{d\H[\PresentP]}{d\delta}$ & $ \mu-\mu\log\mu$ \\[5pt]
\hline\hline
\end{tabular} 
\caption{Information measures and differential information rates of
	continuous-time renewal processes.  See Sec.~\ref{sec:InfoArch} for
	calculational details. Several information measures are omitted as they are
	linear combinations of information measures already in the table; e.g.,
	$\rhomu = \bmu+\qmu$ \cite{Jame11a}.
  }
\label{tab:1}
\end{table*}

\section{Differential information rates}
\label{sec:InfoArch}

We define continuous-time information anatomy \cite{Jame11a} quantities as
\emph{rates}. As mentioned earlier, the present extends over an infinitesimal
time. To define information anatomy rates, we let $\PresentP$ be the symbols
observed over an arbitrarily small length of time $\delta$, starting at the
present $0^-$. It could be that $\PresentP$ encompasses some portion of
$\MT_1$; the notation leaves this ambiguous. The entropy rate is now:
\begin{align}
\hmu = \lim_{\delta\rightarrow 0} \frac{d\H[\PresentP|\Past]}{d\delta}
  ~.
\label{eq:hmuPerCS}
\end{align}
This is equivalent to the more typical random-variable ``block'' definition of entropy rate \cite{Crut01a}:
$\lim_{\delta\rightarrow\infty} \H[\MT_{0:\delta}] / \delta$.
Similarly, we define the \emph{single-measurement entropy rate} as:
\begin{align}
\H_0 = \lim_{\delta\rightarrow 0} \frac{d\H[\PresentP]}{d\delta}
  ~,
\label{eq:H0PerCS}
\end{align}
the \emph{bound information rate} as:
\begin{align}
\bmu = \lim_{\delta\rightarrow 0} \frac{d\I[\FutureP;\PresentP|\Past]}{d\delta}
  ~,
\label{eq:bmuPerCS}
\end{align}
the \emph{ephemeral information rate} as:
\begin{align}
\rmu = \lim_{\delta\rightarrow 0} \frac{d\H[\PresentP|\Past,\FutureP]}{d\delta}
  ~,
\end{align}
and the \emph{co-information rate} as:
\begin{align}
\qmu = \lim_{\delta\rightarrow 0} \frac{d\I[\Past;\PresentP;\FutureP]}{d\delta}
  ~.
\end{align}
In direct analogy to discrete-time process information anatomy, we have the
relationships:
\begin{align*}
\H_0 & = 2\bmu+\rmu+\qmu ~,\\
\hmu & = \bmu+\rmu
  ~.
\end{align*}
So, the entropy rate $\hmu$, the instantaneous rate of information creation,
again decomposes into a component $\bmu$ that represents active information
storage and a component $\rmu$ that represents ``wasted'' information.

\begin{figure}[htp]
\centering
\includegraphics[width=\columnwidth]{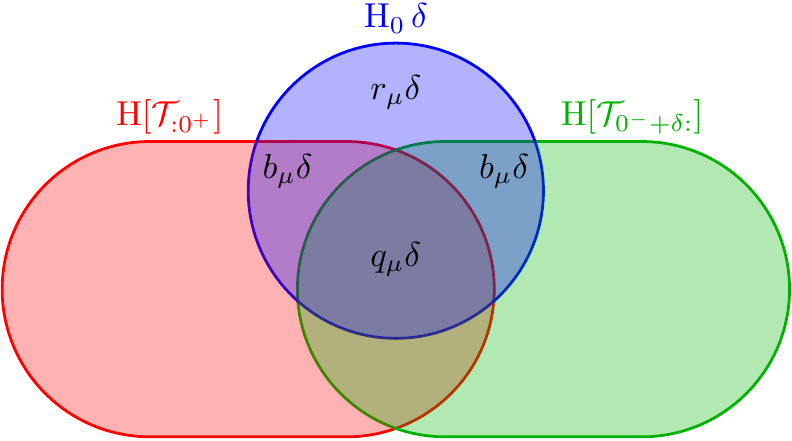}
\caption{Predictively useful and predictively useless randomness for renewal
	processes ($|\ProcessAlphabet| = 1$): Information diagram for the past
	$\Past$, infinitesimal present $\PresentP$, and future $\FutureP$. The
	measurement entropy rate $\H_0$ is the rate of change of the
	single-measurement entropy $\H[\PresentP]$ at $\delta=0$. The ephemeral
	information rate $\rmu =\H[\PresentP|\Past,\FutureP]$ is the rate of change
	of useless information generation at $\delta=0$. The bound information rate
	$\bmu = \I[\PresentP;\FutureP|\Past]$ is the rate of change of active
	information storage. And, the co-information rate $\qmu
	=\I[\Past;\PresentP;\FutureP]$ is the rate of change of shared information
	between past, present, and future. These definitions closely parallel those
	in Ref. \cite{Jame11a}.
  }
\label{fig:Intuition3}
\end{figure}

Prescient states (not necessarily \emph{minimal}) are adequate for deriving all
information measures aside from $\Cmu^{\pm}$.  As such, we focus on the
transition dynamics of noneventually $\Delta$-Poisson \eMs\ and, implicitly,
their bidirectional machines.

To find the joint probability density function of the time to next event $\st^-$ and time since last event $\st^+$, we note that $\st^+ + \st^-$ is an interevent interval; hence:
\begin{align*}
\rho(\st^+,\st^-) \propto \phi(\st^+ + \st^-)
  ~.
\end{align*}
The normalization factor of this distribution is:
\begin{align*}
Z & = \int_0^{\infty} \int_0^{\infty} \phi(\st^+ + \st^-) d\st^+ d\st^- \\
  & = \int_0^{\infty} \int_{\st^-}^{\infty} \phi(\st^+) d\st^+ d\st^- \\
  & = \int_0^{\infty} \st^- \phi(\st^-) d\st^- \\
  & = \mu^{-1}
  ~.
\end{align*}
So, the joint probability distribution is:
\begin{align*}
\rho(\st^+,\st^-) & = \frac{\phi(\st^+ + \st^-)}{Z} \\
  & = \mu \phi(\st^+ + \st^-)
  ~.
\end{align*}
Equivalently, we could have calculated the conditional probability density
function of time-to-next-event given that it has been at least $\st^+$ since
the last event. This, by similar arguments, is $\phi(\st^+ + \st^-) /
\Phi(\st^+)$. This would have given the same expression for
$\rho(\st^+,\st^-)$.

To find the excess entropy, we merely need calculate \cite{Crut08a,Crut08b}:
\begin{align*}
\EE & = \I[\St^+;\St^-] \\
  & = \int_0^{\infty} \int_0^{\infty} \mu\phi(\st^+,\st^-) \log \frac{\mu \phi(\st^+,\st^-)}{\phi(\st^+)\phi(\st^-)} d\st^+ d\st^-
   ~.
\end{align*}
Algebra not shown here gives:
\begin{align*}
\EE = \int_0^{\infty} \mu \, t \, & \phi(t) \log_2
  \big( \mu \, \phi(t) \big) dt \\
    & \qquad - 2 \int_0^{\infty} \mu \, \Phi(t) \log_2 \big( \mu
	\, \Phi(t) \big) dt
  ~.
\end{align*}
Unsurprisingly \cite{pinsker1960information}, this agrees with the formula
given in Ref. \cite{Marz14e}, which was derived by considering the limit of
infinitesimal time discretization.

Now, we turn to the more technically challenging task of calculating
differential information anatomy rates. Suppose that $\Gamma_{\delta}$ is a
random variable for paths of length $\delta$.  Each path is uniquely specified
by a list of times of events. Let $X_{\delta}$ be a random variable defined by: 
\begin{align*}
X_{\delta} & = \begin{cases}
	0 & \text{No events in } \Gamma_{\delta} \\
	1 & 1 \text{ event in } \Gamma_{\delta} \\
	2 & \geq 2 \text{ events in } \Gamma_{\delta}
	\end{cases}
  ~.
\end{align*}
We first illustrate how to find $\H_0$, since the same technique allows
calculating $\hmu$. We can rewrite the path entropy as:
\begin{align*}
\H[\Gamma_{\delta}] = \H[X_{\delta}] + \H[\Gamma_{\delta}|X_{\delta}]
  ~.
\end{align*}
For renewal processes, when $\mu$ can be defined, we see that:
\begin{align*}
\Prob (X_{\delta} = 0) & = 1-\mu \delta + O(\delta^2)~, \\
\Prob (X_{\delta} = 1) & = \mu \delta + O(\delta^2) ~, \text{ and}\\
\Prob (X_{\delta}=2)   & = O(\delta^2)
  ~.
\end{align*}
Straightforward algebra shows that:
\begin{align*}
\H[X_{\delta}] & = \mu\delta - \mu\delta\log (\mu\delta) + O(\delta^2\log\delta)
  ~.
\end{align*}
We would like to find a similar asymptotic expansion for
$\H[\Gamma_{\delta}|X_{\delta}]$, which can be rewritten as:
\begin{align*}
\H[\Gamma_{\delta}|X_{\delta}]
  & = \Prob (X_{\delta}=0) \H[\Gamma_{\delta}|X_{\delta}=0] \nonumber \\
  & \qquad + \Prob(X_{\delta}=1) \H[\Gamma_{\delta}|X_{\delta}=1] \nonumber \\
  & \qquad + \Prob(X_{\delta}=2) \H[\Gamma_{\delta}|X_{\delta}=2]
  ~.
\end{align*}
First, we notice that $\Gamma_{\delta}$ is deterministic given that
$X_{\delta}=0$---the path of all silence. So, $\H[\Gamma_{\delta}|X_{\delta}=0]
= 0$. Second, we can similarly ignore the term $\Prob(X_{\delta}=2)
\H[\Gamma_{\delta}|X_{\delta}=2]$ since $\Prob(X_{\delta}=2)$ is
$O(\delta^2)$ and, we claim, $\H[\Gamma_{\delta}|X_{\delta}=2]$ is $O(\log\delta)$: 
by standard maximum entropy arguments,
$\H[\Gamma_{\delta}|X_{\delta}=2]$ is at most $\log\delta$, and by noting that trajectories with only one event are a strict subset of trajectories with more than one event but with multiple events arbitrarily close to one another, $\H[\Gamma_{\delta}|X_{\delta}=2]\geq \H[\Gamma_{\delta}|X_{\delta}=1]$ which, by arguments below, is $O(\log\delta)$.
Thus, the term
$\Prob(X_{\delta}=2) \H[\Gamma_{\delta}|X_{\delta}=2]$ is
$O(\delta^2\log\delta)$ at most. Finally, to calculate
$\H[\Gamma_{\delta}|X_{\delta}=1]$,
we note that when $X_{\delta}=1$, paths can be uniquely specified by an
event time, whose probability is $\Pr(\mathcal{T}=t|X_{\delta}=1) \propto
\Phi(t)\Phi(\delta-t)$. A Taylor expansion about $\delta / 2$ shows that
$\Pr(\mathcal{T}=t|X_{\delta}=1) = \frac{1}{\delta} + h(t)$ for some $h(t)$ in
which $\lim_{\delta\rightarrow 0} \delta^3 h(t) = 0$ for all $0\leq t\leq
\delta$. So, overall, we find that:
\begin{align*}
\Pr(\Gamma_{\delta}|X_{\delta}=1) = \frac{1}{\delta} + \delta
\Delta \Pr(\Gamma_{\delta}|X_{\delta}=1)
  ~,
\end{align*}
where $\lim_{\delta\rightarrow 0} \delta^2 \Delta
\Pr(\Gamma_{\delta}|X_{\delta}=1) = 0$ for any $\Gamma_{\delta}$ with at least
one event in the path. The largest corrections to
$\Pr(\Gamma_{\delta}|X_{\delta}=1)$ come from ignoring the paths with two or more
events, rather than from approximating all paths with only one event as equally likely. In sum, we see that:
\begin{align*}
\H[\Gamma_{\delta}|X_{\delta}]
  & = \mu\delta\log\delta + O(\delta^2\log\delta)
  ~.
\end{align*}
Together, these manipulations give:
\begin{align*}
\H[\Gamma_{\delta}]
  & = \mu\delta - \mu\delta\log\mu + O(\delta^2\log\delta)
  ~.
\end{align*}
This then implies:
\begin{align*}
H_0 & = \lim_{\delta\rightarrow 0} \frac{d\H[\Gamma_{\delta}]}{d\delta} \\
  & =\mu-\mu\log\mu
  ~.
\end{align*}
A similar series of arguments helps to calculate $\hmu (\st^+)$ defined in
Eq.~(\ref{eq:hmuPerCS}), where now, $\mu$ is replaced by $\phi(\st^+)
/ \Phi(\st^+)$:
\begin{align}
\hmu (\st^+)
  & = \frac{\phi(\st^+)}{\Phi(\st^+)}
  - \frac{\phi(\st^+)}{\Phi(\st^+)} \log \frac{\phi(\st^+)}{\Phi(\st^+)}
  ~,
\label{eq:hmust}
\end{align}
which gives:
\begin{align*}
\hmu & = \int_0^{\infty} \mu \phi(\st^+) d\st^+ \nonumber \\
  & - \int_0^{\infty} \mu
  \phi(\st^+) \log \frac{\phi(\st^+)}{\Phi(\st^+)} d\st^+
  ~.
\end{align*}
Algebra (namely, integration by parts) not shown here yields the expression:
\begin{align}
\hmu = -\mu \int_0^{\infty} \phi(t)\log \phi(t) dt
  ~.
\label{eq:hmufinal}
\end{align}
As expected, this is the nondivergent component of the expression given in Eq.
($10$) of Ref. \cite{Marz14e} for the $\delta$-entropy rate of renewal
processes. And, it agrees with expressions derived in alternative ways \cite{Gira05a}.

We need slightly different techniques to calculate $\bmu$, as we
no longer need to decompose a path entropy. From Eq.~(\ref{eq:bmuPerCS}), we
have:
\begin{align*}
\bmu (\st^+) = \lim_{\delta\rightarrow 0}
  \frac{d\H[\St^-_{\delta}|\St^+_0 = \st^+]}{d\delta}
  ~.
\end{align*}
Let's develop a short-time $\delta$ asymptotic expansion for $\Prob
(\St^-_{\delta}=\st^-|\St^+_0 = \st^+)$. First, we notice that $\St^+_0
\rightarrow \St^+_{\delta} \rightarrow \St^-_{\delta}$, so that:
\begin{align*}
\Prob &(\St^-_{\delta}=\st^- | \St^+_0 = \st^+) \\
  & = \int_0^{\infty} \Prob (\St^-_{\delta}=\st^-|\St^+_{\delta} = \st')
  \Prob (\St^+_{\delta}=\st'|\St^+_{0} = \st^+) d\st'
  .
\end{align*}
We already can identify:
\begin{align*}
\Prob (\St^-_{\delta}=\st^-|\St^+_{\delta} = \st')
  & = \frac{\phi(\st^- + \st')}{\Phi(\st')}
  ~.
\end{align*}
To understand $\Prob (\St^+_{\delta}=\st'|\St^+_{0} = \st')$, we expand:
\begin{align*}
\Prob (\St^+_{\delta}=\st'|\St^+_{0} = \st^+)
  & = \sum_{x=0}^2
  \Prob (\St^+_{\delta}=\st',X_{\delta}=x|\St^+_{0} = \st^+)
  ~.
\end{align*}
Recall that $\Prob(X_{\delta}=2|\St^+_0=\st^+)$ is $O(\delta^2)$, that:
\begin{align*}
\Prob(\St^+_\delta=\st',X_{\delta}=0|\St^+_0=\st^+)
  & = \frac{\Phi(\st')}{\Phi(\st^+)}\delta(\st'-\delta-\st^+)
  ~,
\end{align*}
and that:
\begin{align*}
\Prob(\St^+_\delta=\st' & , X_{\delta}=1|\St^+_0=\st^+) \\
  & = \begin{cases}
  \frac{\phi(\st^++\delta-\st')}{\Phi(\st^+)}\Phi(\st') & \st'\leq\delta \\
  0 & \st'>\delta
  \end{cases}
  ~.
\end{align*}
Then, straightforward algebra not shown gives:
\begin{align*}
\Prob & (\St^-_{\delta}=\st^- |\St^+_0 = \st^+) \\
  & = \frac{\phi(\st^+ + \st^-)}{\Phi(\st^+)}
  + \frac{\phi'(\st^+ + \st^-)+\phi(\st^-) \phi(\st^+)}{\Phi(\st^+)} \delta
  + O(\delta^2)
  ~.
\end{align*}
This can be used to derive:
\begin{align*}
\bmu (\st^+) = \frac{\phi(\st^+)}{\Phi(\st^+)}
  & \Big(\log \phi(\st^+) - 1 \\
  & -\int_0^{\infty} \phi(\st^-) \log \phi(\st^+ + \st^-) d\st^-\Big)
  ~,
\end{align*}
in nats. When $\phi(t) = \lambda e^{-\lambda t}$, for instance, $\bmu(\st^+) =
0$ for all $\st^+$, confirming in a much more complicated calculation that
Poisson processes really are memoryless.  This allows us to calculate the total
$\bmu$ as:
\begin{align*}
\bmu & = \int_0^{\infty} \mu \Phi(\st^+) \bmu (\st^+) d\st^+ \\
     & = -\mu \Big( 1 + \int_0^{\infty} \int_0^{\infty}
	 \phi(t) \phi(t') \log \phi(t+t') dt dt' \\
     & \qquad\qquad - \int_0^{\infty} \phi(t) \log \phi(t) dt \Big)
  ~,
\end{align*}
in nats. And, from this, we find $\rmu$ using:
\begin{align*}
\rmu & = \hmu - \bmu \\
     & = -\mu \int_0^{\infty} \phi(\st^+) \log \phi(\st^+) d\st^+ \\
     & \qquad + \mu \Big( 1 + \int_0^{\infty} \int_0^{\infty}
	 \phi(\st^+) \phi(\st^-) \log \Phi(\st^+) d\st^+ d\st^- \\
     & \qquad\qquad - \int_0^{\infty} \phi(\st^+) \log \phi(\st^+) d\st^+ \Big) \\
     & = -\mu \Big( 2\int_0^{\infty} \phi(t)\log \phi(t+t') dt -1 \\
     & \qquad\qquad - \int_0^{\infty} \phi(t) \int_0^{\infty}
	 \phi(t') \log \phi(t+t') dt' dt \Big)
	 ~.
\end{align*}
Continuing, we calculate $\qmu$ from:
\begin{align*}
\qmu & = \H_0 - (\hmu + \bmu) \\
     & =-\mu\log \mu - \mu \\
     & + \mu \Big(\int_0^{\infty} \phi(t) \int_0^{\infty}
	 \phi(t')\log \phi(t+t') dt' dt + 1 \Big) \\
     & = \mu \int_0^{\infty} \phi(t) \int_0^{\infty}
	 \phi(t')\log \phi(t') dt' dt -\mu\log \mu
	 ~.
\end{align*}
And, we calculate $\rhomu$ via:
\begin{align*}
\rhomu & = \H_0 - \hmu \\
       & = -\mu \log \mu -\mu
       + \mu \int_0^{\infty} \phi(\st^+) \log \phi(\st^+) d\st^+
	   ~.
\end{align*}
All these quantities are gathered in Table \ref{tab:1}, which gives them in
bits rather than nats.

\section{Unifilar Hidden Semi-Markov Models}
\label{sec:UHSMMs}

The \eMs\ of discrete-time, discrete-symbol processes are well understood and,
as we now appreciate from Secs.~\ref{sec:CausalStates}-\ref{sec:InfoArch}, the
predictive equivalence relation defining them readily applies to
continuous-time renewal processes. This gives the latter's analogous maximally
predictive models: continuous or hybrid discrete-continuous \eMs, when
minimal. Here, we introduce a new class of process generators that are unifilar
versions of Ref. \cite{levinson1986continuously}'s hidden semi-Markov models,
but whose dwell time distributions can take any form. (Note that general
semi-Markov models are a strict subset.) Roughly speaking, they are stateful
renewal processes, but this needs to be clarified. Many of their calculations
reduce to those in Secs.~\ref{sec:CausalStates}-\ref{sec:InfoArch}. When
appropriate, we skip these steps.

\begin{figure}[h]
\includegraphics[width=0.7\columnwidth]{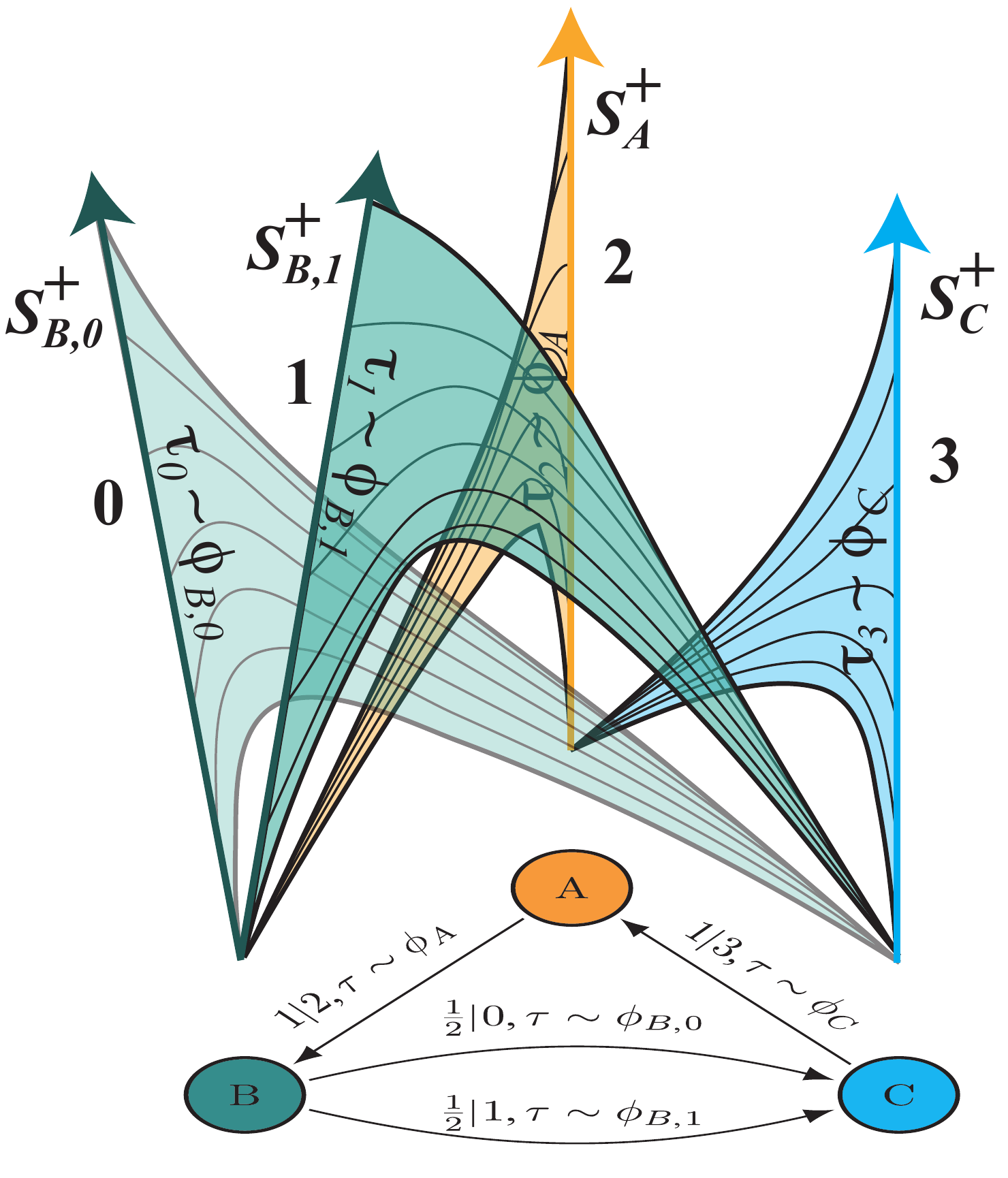}
\caption{Unifilar hidden semi-Markov model: (Top) Prescient machine---an
	\eM\ under mild conditions---for the process emitted by the generator
	below. Causal states $\CS^+_A$, $\CS^+_{B,0}$, $\CS^+_{B,1}$, and $\CS^+_C$
	are isomorphic to $\Re^+$. During an event interval, symbol $\ms \in
	\{0,1,2,3\}$ is emitted. A transition occurs to a new event when the state
	dwell time is exhausted at $\tau_\ms$, which is distributed according to
	$\phi_{\cs,\ms}$. (Bottom) Generative model with three hidden states ($A$,
	$B$, and $C$) emits symbols $\ms \in \{0, 1, 2, 3\}$ for dwell times $\tau$
	drawn from probability density functions $\phi_A$,
	$\phi_{B,0}$,$\phi_{B,1}$, and $\phi_{C}$, respectively.  (Transition
	labels as in Fig. \ref{fig:genModel}.)
	}
\label{fig:UHSMM}
\end{figure}

We start by introducing the minimal generative models in Fig.~\ref{fig:UHSMM}.
Let $\Gen$ be the set of states in this generative model. Each state $g\in\Gen$
emits a symbol $\ms\in\MeasAlphabet$ and a dwell time $\tau \sim \phi_{g,\ms}$
for that symbol, and, based on the state $g$ and emitted symbol $\ms$,
transitions to a new state $g'$. We assume that the underlying generative model
is \emph{unifilar}: that the new state $g$ is uniquely specified by the prior
state $g$ and emitted symbol $\ms$. We introduce a perhaps unfamiliar
restriction on the labeled transition matrices $\{T_g^{(\ms)}: \ms \in
\MeasAlphabet \}$. Define $\text{supp}(g) := \{\ms:p(\ms|g)>0\}$ and
$\text{supp}(g\rightarrow g') := \{\ms:p(g',\ms|g)>0\}$. Then, we focus only on
generative models for which $\text{supp}(g)\cap \text{supp}(g\rightarrow g') =
\emptyset$. This simply ensures that there is no uncertainty in when one dwell
time finishes and another begins. For example, consider the generator in
Fig.~\ref{fig:UHSMM}(bottom): if states $B$ and $C$ were both to emit a $0$ in
succession, it would be impossible to tease apart when the process switched
from state $B$ to state $C$. The restriction introduces no loss of generality
for our purposes.


A prescient model of this combined process is shown in
Fig.~\ref{fig:UHSMM}(top). Each state $g\in\Gen$ comes equipped with one or
more renewal process-like tails (semi-infinite spaces that act as continuous
counters) that generically take the form of Fig.~\ref{fig:NEDP}. The leakiness
of these (dissipative) counters is given by $\phi_{g,\ms}$, the probability
density function from which the dwell time is drawn. This new form of
state-transition diagram depicts the \eM\ of these hidden semi-Markov
processes. Moreover, if one or more of the dwell-time distributions gives an
eventually Poisson or an eventually-$\Delta$ Poisson structure, the
presentation in Fig.~\ref{fig:UHSMM}(top) is a prescient machine, but not the \eM.

More generally, any such unifilar minimal generative model has a prescient
machine with a ``node'' for each underlying hidden state $g$ and as many
counters as needed---one for every almost-everywhere unique $\phi_{g,\cdot}$.
Each counter leaks probability to the next underlying hidden state $g'$, which
is completely determined by $g$ and $\ms$.

{\The The presentation in Fig. \ref{fig:UHSMM}(top) is a prescient machine
for the process generated by the unifilar hidden semi-Markov model of Fig.
\ref{fig:UHSMM}(bottom).
\label{the:2}
}

{\ProThe To show that this is a prescient machine, we need to show that the
present model state-- consisting of hidden state $g$, current emitted symbol
$x$, and dwell time $\tau$-- is uniquely specified by the observed past almost
surely. The observed symbol $\ms$ is given by the current symbol in the
observed past. The restriction on successive emitted symbols (that
$\text{supp}(g)\cap \text{supp}(g\rightarrow g') = \emptyset$) implies that the
observed dwell time $\tau$ is exactly the observed length of $\ms$.  Finally,
the underlying hidden state $g$ is determined uniquely by a \emph{function} of
the past almost surely, in which all dwell-time information is removed, by
assumption: the restriction mentioned earlier implies there is no uncertainty
in when one dwell time finishes and another begins. And, the unifilarity of the
dynamic on hidden states $g$ implies that the sequence of symbols in the
observed past are sufficient to specify the hidden state $g$ almost surely.
Hence, $g$ is determined uniquely from the observed past almost surely. The
theorem follows.
}

{\Rem Theorem~\ref{the:2} can be straightforwardly generalized to specify
conditions under which the presentation is an \eM, a minimal prescient machine,
by incorporating the conditions of Thm.~\ref{the:RenewCausalStates}.}

The stationary distribution for $\rho(\tau|g,\ms)$ directly follows the
treatment for the continuous-time renewal processes in Sec.~\ref{sec:NEDP}, and
so:
\begin{align*}
\rho(\tau|g,\ms) = \mu_{g,\ms}\Phi_{g,\ms}(\tau)
  ~,
\end{align*}
where $\mu_{g,\ms} = 1/\int_{0}^{\infty}\Phi_{g,\ms}(\tau) d\tau$. Then we note
that:
\begin{align*}
p(\ms|g) & \propto \frac{T^{(\ms)}_g}{\mu_{g,\ms}} \\
   \rightarrow p(\ms|g)
   & = \frac{T^{(\ms)}_g/\mu_{g,\ms}}{\sum_{\ms'} T^{(\ms')}_g/\mu_{g,\ms'}}
  ~.
\end{align*}
And so:
\begin{align}
\rho(g,\ms,\tau)
  & = p(g) \frac{T^{(\ms)}_g/\mu_{g,\ms}}{\sum_{\ms'} T^{(\ms')}_g/\mu_{g,\ms'}}
  \mu_{g,\ms} \Phi_{g,\ms}(\tau) \nonumber \\
  & = p(g) \frac{T^{(\ms)}_g}{\sum_{\ms'} T^{(\ms')}_g/\mu_{g,\ms'}}
  \Phi_{g,\ms}(\tau)
  ~.
\label{eq:foo}
\end{align}
To find $p(g)$, we again calculate the probability mass dumped at $\tau=0$ in terms of $p(g)$:
\begin{align*}
p(g,\ms,0)
 & = \sum_{g',\ms'}\int_0^{\infty} p(g',\ms',\tau) \delta_{\epsilon(g',\ms'),g}
 \frac{\phi_{g',\ms'}(\tau)}{\Phi_{g',\ms'}(\tau)}  T^{(\ms)}_{g} d\tau
  ~.
\end{align*}
After a straightforward substitution of Eq.~(\ref{eq:foo}) and noting that
$\int_0^{\infty} \phi_{g,\ms}(\tau) = 1$, we find:
\begin{align*}
p(g) \frac{1}{\sum_{\ms'} T^{(\ms')}_g/\mu_{g,\ms'}}
  & = \sum_{g',\ms'}
  \frac{1}{\sum_{\ms} T^{(\ms)}_g/\mu_{g,\ms}} T^{(\ms')}_{g',g} p(g')
  ~.
\end{align*}
So: 
\begin{align*}
  \frac{p(g)}{\sum_{\ms} T^{(\ms)}_g/\mu_{g,\ms}}
  & = \sum_{g'} T_{g',g} \frac{p(g')}{\sum_{\ms} T^{(\ms)}_g/\mu_{g,\ms}}
  ~.
\end{align*}
Let $\pi(g)$ be the stationary distribution for the underlying discrete-state \eM: 
\begin{align*}
\pi := \text{eig}_1 (T)
  ~,
\end{align*}
where the eigenvector is normalized such that the sum of its entries is $1$.
Then:
\begin{align*}
\pi(g) & \propto \frac{p(g)}{\sum_{\ms} T^{(\ms)}_g/\mu_{g,\ms}}
  ~.
\end{align*}
Or, rewriting and normalizing, we have:
\begin{align*}
  p(g) & = \pi(g)
  \frac{\sum_{\ms} T^{(\ms)}_g/\mu_{g,\ms}}{\sum_{g',\ms} \pi(g') T^{(\ms)}_{g'}/\mu_{g',\ms}}
  ~.
\end{align*}
Altogether, we find that the steady-state distribution is given by:
\begin{align}
\rho(g,\ms,\tau)
  & = \left(\pi(g) \frac{\sum_{\ms'} T^{(\ms')}_g/\mu_{g,\ms'}}{\sum_{g',\ms'} \pi(g') T^{(\ms')}_{g'}/\mu_{g',\ms'}}\right)
  \nonumber \\
  & \qquad \times \left(\frac{T^{(\ms)}_g/\mu_{g,\ms}}{\sum_{\ms'} T^{(\ms')}_g/\mu_{g,\ms'}}\right)
  \mu_{g,\ms} \Phi_{g,\ms}(\tau) \nonumber \\
  & = \frac{\pi(g) T^{(\ms)}_g \Phi_{g,\ms}(\tau)}{\sum_{g',\ms'} \pi(g') T^{(\ms')}_{g'}/\mu_{g',\ms'}}
  ~.
\label{eq:steadystate}
\end{align}
Using the formulae for entropies of mixed random variables
\cite{Nair06a}, we find a statistical complexity of:
\begin{align*}
\Cmu^+ & = \H[\rho(g,\ms,\tau)] \\
       & = \left\langle \H[\rho(\tau|g,\ms)] \right\rangle_{p(g,\ms)}
	   + \H[p(g,\ms)] \\
       & = \left\langle \int_0^{\infty} \mu_{g,\ms} \Phi_{g,\ms}(\tau)
	   \log \frac{1}{\mu_{g,\ms}\Phi_{g,\ms}(\tau)} d\tau
	   \right\rangle_{p(g,\ms)} \\
       & \qquad + \H \left[ \frac{\pi(g)
	   T_g^{(\ms)}/\mu_{g,\ms}}{\sum_{g',\ms'} \pi(g')
	   T_{g'}^{(\ms')}/\mu_{g',\ms'}} \right]
	   ~.
\end{align*}
Note that $\H[\pi(g)]$ is the statistical complexity of the underlying
discrete-time \eM\ and that $\H[\rho(\tau|g,\ms)]$ is the statistical
complexity of a noneventually $\Delta$-Poisson renewal process with interevent
distribution $\phi_{g,\ms}(\tau)$, averaged over $g$ and $\ms$. Hence, the
statistical complexity of these unifilar hidden semi-Markov processes differs
from the statistical complexity of its ``components'' by:
\begin{align*}
\H \left[\frac{\pi(g) T_g^{(\ms)}/\mu_{g,\ms}}{\sum_{g',\ms'} \pi(g')
T_{g'}^{(\ms')}/\mu_{g',\ms'}} \right]
  - \H[\pi(g)]
  ~.
\end{align*}
Whether this difference is positive or negative depends on both matrices
$T_g^{(\ms)}/\mu_{g,\ms}$. In general, we expect the difference to be
positive.

Since there are multiple observed symbols $\ms \in \ProcessAlphabet$ generated
by these machines, $\H_0$ (and so $\rhomu$) and $\bmu$ as defined in
Sec.~\ref{sec:InfoArch} diverge. However, the entropy rate $\hmu$ and excess
entropy $\EE$ as defined in Sec.~\ref{sec:InfoArch} do not diverge for
processes generated by this restricted class of unifilar hidden semi-Markov
models. From the steady-state distribution given in Eq.~(\ref{eq:steadystate})
and from the entropy rate expressions in
Eqs.~(\ref{eq:hmust})-(\ref{eq:hmufinal}) of Sec.~\ref{sec:InfoArch}, we
immediately have the entropy rate for these unifilar hidden semi-Markov models:
\begin{align}
\hmu & = \lim_{\delta\rightarrow 0}
  \frac{d \H[\Gamma_{\delta}|\St^+_{0^+} = (g,\ms,\tau)]}{d\delta} \nonumber \\
  & = \sum_{g,\ms} \rho(g,\ms)
  \left(-\mu_{g,\ms} \int_0^{\infty} \phi_{g,\ms}(\tau)\log \phi_{g,\ms}(\tau) d\tau\right)
  \nonumber \\
  & = -\frac{\sum_{g,\ms} \pi(g) T_g^{(\ms)} \int_0^{\infty} \phi_{g,\ms}(\tau)\log \phi_{g,\ms}(\tau) d\tau}
  {\sum_{g',\ms'} \pi(g') T_{g'}^{(\ms')}/\mu_{g',\ms'}}
  ~.
\label{eq:hmu_HSMM}
\end{align}
To ground intuition, recall that each state in the underlying \eM\ for
semi-Markov processes corresponds to a unique observation symbol. Hence,
setting $\pi$ to $\text{eig}_1 (T_{g,g'})$ and noting that each $g$ is uniquely
associated to some $\ms$ in Eq.~(\ref{eq:hmu_HSMM}) recovers the results of
Ref. \cite{Gira03} for the entropy rate of semi-Markov processes, though the
notation differs somewhat \footnote{Actually, we would apply
Eq.~(\ref{eq:hmu_HSMM}) to a semi-Markov process in reverse-time so that the
underlying model is unifilar rather than co-unifilar; but entropy rate is
invariant to time reversal \cite{Crut01a}.}.

The process' excess entropy $\EE = \I[\CS^+;\CS^-]$ can be calculated if we can
find the joint probability distribution $\Pr(\st^+,\st^-)$ of forward- and
reverse-time causal states. To this end, we add an additional restriction on
the generative model: we focus only on generative models for which
$\text{supp}(g')\cap \text{supp}(g\rightarrow g') = \emptyset$. With this
restriction on labeled transition matrices, the time-reversed \eM\ of the
process has the same form as the \eM\ of the forward-time process, but with a
different $\Gen$. The latter is related to the forward-time $\Gen$ via
manipulations described in Ref. \cite{Crut08b}. As such, we can
write down $p(\st^+,\st^-)$:
\begin{align*}
\Pr(\st^+ = (\tau,g,\ms) & | \st^- = (g',\ms',\tau')) \\
  & = \Pr(\tau|g,\ms,\tau')\delta_{\ms,\ms'} \Pr(g|g',\ms',\ms,\tau') \\
  & = \phi_{g,\ms}(\tau+\tau') \Pr(g|g',\ms)\delta_{\ms,\ms'}
  ~,
\end{align*}
where we obtain $p(g|g',\ms)$ from standard methods
\cite{Crut08b,Elli11a} applied to (only) the dynamic on $\Gen$.
Note that $p(g|g',\ms',\ms,\tau')$ reduces to $p(g|g',\ms')$ as $g'$ and $\ms'$
uniquely specify the distribution from which $\tau'$ is drawn and since
$\ms=\ms'$. We leave the the final steps to $\EE$ as an exercise.


\section{Conclusions}
\label{sec:Apps}

Though the definition of continuous-time causal states parallels that for
discrete-time causal states, continuous-time \eMs\ and information measures are
markedly different from their discrete-time counterparts. Similar technical
difficulties arise more generally when describing minimal maximally predictive
models of other continuous-time, discrete-symbol processes that are not the
continuous-time Markov processes analyzed in Ref. \cite{Riec16a}. The resulting
\eMs\ do not appear like conventional HMMs---recall
Figs.~\ref{fig:NEDP}-\ref{fig:EDP} and, especially, Fig.
\ref{fig:UHSMM}(top)---and most of the information measures---excepting the
excess entropy---are reinterpreted as differential information rates.

Moreover, the \eM\ continuous-time machinery gave us a new way to calculate
these information measures. Traditionally, expressions for such information
measures come from calculating the time-normalized path entropy of arbitrarily
long trajectories; e.g., as in Ref. \cite{Gira03}. Instead, we calculated the
path entropy of arbitrarily short trajectories, conditioned on the past. This
allowed us to extend the results of Ref. \cite{Gira03} for the entropy rate of
continuous-time discrete-output processes to a previously untouched class of
processes---unifilar hidden semi-Markov processes.

There are two immediate practical benefits to an in-depth look at the \eMs\ of
continuous-time hidden semi-Markov processes. First, statistical model
selection when searching through unifilar hidden Markov models is significantly
easier than when searching through nonunifilar Hidden Markov models
\cite{Stre13a}, and these benefits should carry over to the case of
continuous-time \eMs. Second, the formulae in Table \ref{tab:1} and those in
Sec.~\ref{sec:UHSMMs} provide new approaches to binless plug-in information
measure estimation; e.g., following Ref. \cite{victor2002binless}.

The machinery required to use continuous-time \eMs\ is significantly different
than that accompanying the study of discrete-time \eMs. Our results here pave
the way toward understanding the difficulties that lie ahead when studying the
structure and information in continuous-time processes.


\section*{Acknowledgments}

The authors thank the Santa Fe Institute for its hospitality during visits.
JPC is an SFI External Faculty member. This material is based upon work
supported by, or in part by, the U. S. Army Research Laboratory and the U. S.
Army Research Office under contract number W911NF-13-1-0390. SM was funded by a
National Science Foundation Graduate Student Research Fellowship, a U.C.
Berkeley Chancellor's Fellowship, and the MIT Physics of Living Systems
Fellowship.



\end{document}